# Towards infrared toroidal photodevices: A Review

Burak Gerislioglu,[1][†] and Arash Ahmadivand[2][*][†]

[1]Department of Physics & Astronomy, Rice University, 6100 Main St, Houston, Texas 77005, United States

[2]Department of Electrical & Computer Engineering, Rice University, 6100 Main St, Houston, Texas 77005, United States

*aahmadiv@rice.edu

[†]Equal contribution

**Abstract-** Plasmon excitations in metallic nanostructures can decay directly into dynamic electron-hole pairs (EHPs), exploitable for photocurrent generation. This approach has extensively been employed to develop nanoplasmonic light-sensing devices with significant responsivity and quantum efficiency. Of particular interest is infrared plasmonic photodetectors with a wide range of technological applications, including spectroscopy, biosensing, and surveillance. This Review discusses fundamentals, recent advances, and trending mechanisms in the understanding and applications of plasmon-enhanced photocurrent generation in nanostructures across the infrared spectrum. By highlighting and comparing the developed techniques, we demonstrate the newly introduced directions toward achieving high-photon yield infrared plasmonic photodetection tools. As a promising concept in modern photonics, we represent the emergence of toroidal meta-atoms as plasmon-induced carrier generators with unconventionally exquisite properties for designing advanced, rapid, and next-generation plasmonic photodetectors with significantly high responsivity and photocurrent.

**Keywords-***Plasmonics; infrared photodetectors; toroidal photodetectors; plasmon-induced carrier generation.*

# 1. Introduction

The conversion of incident photons from free space into electrical signals was considered and explained originally as an electrostatic principle, introduced by Hertz while analyzing the influence of the ultraviolet beam on the electric discharge from the conductive probes.[1,2] Thereafter, scientists extensively developed this effect *via* considering and studying the light as a set of discrete wavepackets with a quantized energy, defining by Planck-Einstein energy-frequency relation.[3-5] Such a theory allowed to comprehend the feasibility of the excitation and ejection of electrons from metallic components using beam with particular energy and intensity.[6,7] All these accomplishments initiated and paved the required fundamental methods toward tailoring devices for photocurrent generation by several sophisticated electrical, physical, chemical, mechanical, and thermal procedures. In all elucidated mechanisms, the formation of electron-holes pairs (EHPs) with ultrashort lifetimes (around a few nano- or microseconds) is the direct result of the excitation and decay of electron from subwavelength metallic structures. In the active regime, sweeping of electrons and holes by applying forward and reverse biases, respectively, gives rise to high-photon yield and significant photocurrent, which are important parameters in defining the quality of photodetection systems.[7-10]

In terms of the optical physics, the interaction of an intense light with metallic particles in subwavelength limits leads to the coherent oscillations of free electrons at the d-band in noble metals, known as plasmons,[7,11-15] were perceived in the field of surface sciences by Ritchie *et al.*[16] in the past century. Such resonant interaction between the electromagnetic field and the surface charge oscillation constitutes the collective plasmon oscillations, which enabled versatile properties for a wide range of commercial applications including but not limited to light harvesters,[17-19] chip-scale immunobiosensors,[20-23] modern healthcare tools,[24,25] drug delivery,[26,27] nano-imaging,[28,29] superlensing and superfocusing,[28,30,31] lasers,[32-35] metamaterials,[36-40] quantum technologies,[41-43] and optoelectronic devices.[44-49]

Considering the plasmonic-enhanced optoelectronics devices, the ever-increasing demand for higher responsivity and faster operation speed drives efforts toward exploiting plasmonic nanostructures in designing optoelectronic devices such as *phototransistors* and *photodetectors*.[50] For the latter example, promisingly, plasmonics has remarkably revitalized the spectral and electrical properties of photodetection platforms from the ultraviolet to the far-infrared and terahertz frequencies.[51-56] Among these devices, strategically, infrared nanoplasmonic light sensors possess a wide range of important applications including but not limited to missile warning, target recognition, night and machine vision, astronomy, modern biotechnology, and low-power wavelength division multiplexing (WDM) for short-distance optical communication.[57-60] So far, several efficient approaches have been carried out to amplify the performance, quantum efficiency, and speed of infrared plasmonic photodetectors such as 1) introducing 2D materials (e.g. graphene, $MoS_2$) to the metallic systems,[61-65] 2) quantum dots-mediated plasmonic systems,[66-70] 3) using plasmon-induced hot carrier nanosystems,[7,10,64,71,72] 4) combining generated carriers from hot electron excitation and free-carrier absorption (FCA) principles,[56] and 5) utilizing unconventional spectral features with giant electromagnetic field confinement.[73]

In this focused Review, we illustrate and highlight the recent advances and state-of-the-art technologies toward developing high-responsive nanometer-scale infrared plasmonic photodetectors. By providing an overview of the plasmonic infrared photodetectors, we indicate that why they are being considered for implementing advanced metaphotonics tools. We examine the techniques that have been used to optimize the characteristic response of infrared beam sensing devices. In addition, by discussing and comparing the spectral and electrical properties of several types of near-infrared (NIR) and infrared photodetection platforms, we conclude the context by emphasizing the newly introduced mechanisms which can change the route of the conventional photocurrent generator structures.

## 2. Underlying principles in plasmon-enhanced photocurrent generation

The plasmon-induced carrier generation fundamentally depends on the optical absorption, which can be defined quantitatively *via* integrating the product of wavelength/frequency, local electric-field strength, and the imaginary part of the dielectric permittivity (Im($\varepsilon$)) across the entire volume of the subwavelength metallic structure.[10,74] Although plasmonic nanostructures exhibits strong absorption cross-section, the radiative and nonradiative damping, back-scattering, and inelastic collisions drastically reduce the injection of energetic electrons into the substrate. On the other hand, the generated carriers from the nanoparticle surface can be affected by scattering, recombination, and losing the energy during surmounting energy barriers. Hence, carriers closer to the interface than the *mean-free path* can contribute in the generation of photocurrent, where the relevant electron transfer number can be determined by solving $\frac{1}{2}\int |\mathbf{E}(\mathbf{r})|^2 \text{Im}(\varepsilon) d\mathbf{r}\omega$ over the volume within a specific distance from the dynamic interface.[55,75,76] Differentiating the type of generated carriers plays fundamental role in estimating the efficiency of the developed device, however, the introduced method above fails to address this. In plasmonic platforms, to understand the origin of the generated EHPs, one needs to distinguish the photoexcited and plasmon-decay carriers. Conventionally, two types of interfaces determine the operating mechanism of a photodetection device: 1) Schottky junction and 2) Ohmic junction. For the first interface, a Schottky barrier forms at a metal-semiconductor interface and only passes current flow in one direction by preventing rapid recombination of carriers and collecting either the electrons or holes. Conversely, the Ohmic junction does not hold any effective barrier height and the generated photocurrent contains the participation of interband electrons as well.

As demonstrated in Figure 1a, Zheng *et al*.[76] demonstrated that in the Schottky contact limit, the plasmon-induced carrier merely incorporate the photocurrent generation process. More precisely, for the gold-semiconductor ($TiO_2$) interface, the photogenerated carriers majorly stem from the metal d-band below the Fermi energy level (~2.3 eV).[77,78] In this type of interfaces, the plasmon-induced carrier generation acts as a dominant parameter, which allows for the excitation of electrons from near the Fermi energy level, resulting in significantly dynamic and higher-energy electrons.[76,79-81] Consequently, the source of the photocurrent is only from the plasmon decay. On the other hand, for the Ohmic contact regime, in the absence of barriers, the photocurrent will be generated from the plasmon-decay and low energy interband electrons (see Figure 1b). Here, the thin titanium interlayer between the gold and semiconductor was use to enforce the Fermi energy level of gold to align with the conduction band of semiconductor substrate. The difference between two types of contacts can be better understood by comparing the characteristic current-voltage (*I-V*) graphs, as depicted in Figures 1c and 1d. As can be seen in the panels, the Schottky device shows current enhancement with an exponential lineshape, while the Ohmic device reflects linear *I-V* characteristics. The provided comparison by Zheng and colleagues verifies that the measured photocurrent in the Ohmic device is in agreement with the electron injection, and not elucidated by variations in the device conductance or junction resistances.

Furthermore, this study theoretically demonstrated that the plasmon-induced carrier generation in Schottky interface devices is independent of interband carrier generation. In order to distinguish these type of contributions, the provided integral above must be revised. This is because the developed integration involves the absorption response from interband transitions. Thus to accurately model the participation of electrons in the photocurrent generation, one should employ the following integration over the volume ($V_{MFP}$) of the *mean-free path* to define the field-intensity enhancement: $E_{MFP}^2 \frac{1}{V_{MFP}} \int |\mathbf{E}(\mathbf{r})|^2 d\mathbf{r}$. To ensure that the plasmon-enhanced carrier generation occurs independently of material absorption, using experimental measurements, the researchers demonstrated that the photocurrent response matches with the E-field intensity enhancement rather than with the material-dependent absorption. Relatively, it is well-accepted that increasing the field-enhancement near the interface is important for increasing the efficiency of plasmon-enhanced nanodevices.[82-84]

## 3. Hot electron-based photodetection

Photovoltaic tools based on photoexcitation and immediate charge separation within EHPs have been exceedingly studied over the last two decade.[7,85,86] The popularity of this concept is diminished gradually, due to its limited capability of detecting photons with energies lower than the semiconductor bandgap. To address this concern, the energy of photo-ejected electrons, known as *energetic or hot electrons*, from a metal can be harvested [More information about the internal photoemission process from a metal can be found in Ref. 8]. To this end, one can utilize planar metal-oxide-metal (MOM) structures as the most basic energy harvesting platform by applying photon energy higher than the oxide barrier energy. This methodology has been utilized since 1970 to develop mid-infrared and terahertz photodetectors.[87,88] In both mechanisms, the photoelectrons ejected from the metal cannot surmount the energy barrier (~1 eV), occurred at the metal/oxide interface. Thus, the thickness of the oxide layer can be reduced down to ~1 nm to make the electron tunneling possible, but still there are many deficiencies (e.g. low photocurrent, inefficient coupling of the illumination energy to the structure) that hinders the use of this design modality towards high operating frequency (e.g. near-infrared or visible) for practical applications.[89] Owing to the arguments above, the field tends to find alternative pathways to enhance the hot electron generation and its subsequent emission across the barrier, and surface plasmon excitations in judiciously engineered metallic antenna geometries is a solution to increase the power conversion efficiency and photocurrent response, by controlling the electric field (E-field) distribution along the system.[90-93]

**3.1 Plasmonic grating antenna-based hot electron photodetection**

Periodically aligned slits in a metal film, known as gratings, can strongly couple the impinging electromagnetic wave to the plasmons of the devised metallic nanoplatform, which gives rise to an intense and narrow-band absorption, and highly localized electromagnetic fields. Taking advantage of this approach, for the first time, Sobhani *et al.*[94] demonstrated a grating-based hot electron device that provides significantly large photocurrent responsivity and internal quantum efficiency (IQE) in comparison to previous studies.[74] Similar to earlier plasmonic light sensors, the responsivity of the proposed system is 0.6 mAW$^{-1}$ without applying any bias voltage. Besides, the IQE of the device is around 0.2%, which is 20 times larger than other nanoantenna-based devices (~0.01%),[90] and the responsivity of the system can preserves its narrowband characteristics (FWHM ~54 meV). With the help of its simple grating geometry, one can tune the maximum responsivity of the tool from 1295 nm to 1635 nm. Figure 2a illustrates an artistic illustration of the proposed device with the corresponding geometrical parameters. It is important to note that a 2 nm thick titanium layer affects the spectral response of the nanosystem, which is necessary to adhere the gold gratings on top of the silicon substrate and to adjust the Schottky barrier height. Additionally, a top-view SEM image of the fabricated device is presented in Figure 2b, in which the gold layer thickness (*T*) is varied to determine its impact on the device properties. For all considered thickness values (200 nm, 170 nm, and 93 nm), a single and narrowband responsivity peak is observed, where the width of each peak is independent from the grating thickness. As clearly seen from Figure 2c, the peak responsivity is increased by a factor of seven when the thickness of the gold layer is altered from 93 nm to 200 nm. This significant change indicates the dependence of the absorption cross-section on the grating thickness, which enables a facile tailoring of the responsivity feature. The physical mechanism behind this process is explained in Figure 2d. As plotted in the graph, the thickness dependence of the absorption is because of the constructive or destructive interference between surface plasmons at the upper and lower surfaces of each gold pitch. By carefully selecting the grating thickness, one can minimize the destructive interference, which maximize the plasmon decay and amplify the induced photocurrent.[95] Further investigations showed that the localization of absorption losses contributes to the high responsivity feature of the grating-Schottky device. As it was mentioned in the previous section of this article, generally, the mean free path of the hot electrons in gold is around 35 nm,[96] which means that one should design the grating geometry to effectively use the formed hot electrons near the metal-semiconductor interface. Otherwise, the energy of the hot electron remains below the barrier height and will be reflected back into to the metal without contributing to the photocurrent. To this end, the design was renewed in order to be able to absorb almost all the incident wave at the Schottky interface (see Figure 2e). In continue, the authors investigate the photocurrent spectra of the grating-Schottky tool for the following parameters: *T*=200 nm, interslit distance (*D*)=1100 nm, slit width (*W*)=250 nm. As illustrated in Figure 2f, the spectrum indicates two peaks at 1635 nm and 1285 nm, which are superimposed by following labels 'Mode i' and 'Mode ii', respectively. For each mode, the associated surface charge distribution were provided, where each red and blue area at the interface symbolizes an anti-node of charge accumulation (see Figures 2g and 2h). Particularly, for the 'Mode i', the distribution shows five anti-nodes, the 'Mode ii' indicates six anti-nodes, and the total number of anti-nodes describe the mode number of the surface plasmon.

In the work presented by Li *et al.*,[97] an ultracompact circularly polarized beam detector that integrates well-engineered chirality with hot electron injection is experimentally demonstrated. As a key parameter, the structured photodetector can differentiate right and left circularly polarized light without requiring any additional optical element. In Figure 3a, the proposed 'Z' shaped silver chiral metamaterial on top of a polymethyl methacrylate (PMMA) spacer is illustrated. As depicted in Figure 3b, the standalone shape of the provided design allows to create nanowires, in which a silver bus bar is utilized to electrically connect those nanowires. Next, to be able to form a Schottky barrier, the authors placed an n-type silicon wafer on top of the metallic grating layer (see Figure 3c). In this configuration, the incident light is transmitted along the silicon layer, in which the specific photons (depending on their handedness) are absorbed by the chiral metamolecule. As a result of this process, electrons moved to the higher energy states (as hot electrons), where they can emit over the Schottky barrier to form an electric current. Further, the authors investigate the detection performance of the developed subwavelength device. They fabricated both right handed (RH) and left handed (LH) arrays (see Figures 3d and 3e) to check the effect of high circular dichroism (CD) on the circularly polarized light (CPL) photodetector. It is significant to note that a double-side polished n-type silicon wafer is used to cancel any possible scattering from the front side of the wafer. Besides, aluminum-doped silver is preferred to minimize the silver film roughness and to expedite the formation of an alumina ($Al_2O_3$) layer to preserve the silver film from corrosion.[98] In Figure 3g, the LH metamaterial shows a near unity experimentally measured absorption of the LCP light at 1340 nm, while then RCP light is mostly reflected. Conversely, the RH metamaterial demonstrates almost opposite response, as expected (see Figure 3h). One can observe that the off-resonant absorption of the high, which is due to the additional Titanium adhesion layer and aluminum doping of silver. Nevertheless, the proposed design still has a CD of 0.72 at 1340 nm (see Figure 3i). Following, the researchers examined the photoresponsivity of the device. As pointed out in Figures 3j and 3k, each photoresponsivity spectrum is consistent with the corresponding absorption spectra. The maximum photoresponsivity reaches 2.2 mAW$^{-1}$, with a quantum yield of 0.2%. These values are similar to other Schottky interface-based photodetectors in this operation bandwidth.[91,94] Besides, the experimental measured and theoretically calculated photoresponsivity spectrum are in a good agreement, except the lineshape broadening in the experimental results. Moreover, the large CD provides a substantial polarization discrimination ratio (~3.4) for the induced photocurrent for LCP and RCP light conditions, in which the difference reaches up to 1.5 mAW$^{-1}$. This clearly verifies the ability of the proposed device to distinguish LCP and RCP lights within a compact geometry. Lastly, as a proof of principle, the authors combined multiple RH and LH metamolecules (or enantiomers) into a single platform to create a multifunctional photodetector (in the form of Vanderbilt University logo (see Figure 3m)). In this Figure, the black and white regions of the logo are filled with LH and RH enantiomers, respectively. As predicted, in Figures 3o and 3p, the logo possesses a distinct contrast in reflection under LCP and RCP light, whereas an image of the logo does not come out under un-polarized or linearly polarized light conditions. As a final test, the performance of this platform is explored through scanning photocurrent measurements using focused CP laser illumination. Correspondingly, a good contrast between each photocurrent scanning maps is obtained (see Figures 3q and 3r).

## 3.2 Graphene-based hot electron photodetection

In addition to hot electron generation *via* plasmonic grating antennas, graphene-based platforms have also been studied copiously as a notable candidate for photodetection,[99-101] photovoltaics,[102] and terahertz modulators,[103-105] owing to graphene's high Fermi velocity of charge carriers, extraordinary room-temperature mobility, and atomic layer thickness.[106-109] To this end, Du *et al*.[110] investigated the tunneling effect of hot electrons in a hybrid gold-graphene nanoplatform. In the proposed device design, the hot electrons produced by the gold nanoparticles (NPs) are vertically tunneled through the graphene monolayer, where it behaves as a barrier between the top and bottom electrodes. As a result of this process, a strong vertical photocurrent is measured, where its intensity reaches a maximum value at the resonant wavelength. It is also demonstrated that this photocurrent can be tuned *via* the bias voltage between the top and bottom electrodes, and the incident laser power and wavelength. In Figure 4a, an illustration for the proposed graphene-based photodetector device is provided. The platform consists of a monolayer graphene sheet sandwiched between each gold electrodes, randomly deposited gold NPs on the graphene layer, and silicon/glass (SiO$_2$) substrate. To confirm the structure of the pristine graphene sheet, a similar device without including gold NPs is prepared, and the corresponding optical microscope image is presented in Figure 4b. Unlike the other typical tunneling barriers (e.g. Al$_2$O$_3$ and magnesium oxide (MgO)), a highly uniform single-atom thick graphene barrier allows a full control over the characteristics of the barrier. For instance, the authors tailored the thickness of the barrier by transferring additional graphene monolayers to create disordered stacking. It should be underlined that the strong interaction between each layer is prevented by using PMMA during the transfer procedure. In Figure 4c, the precise thicknesses of the monolayer and bilayer graphene sheets are depicted, which are measured using tapping-mode atomic force microscopy (AFM), and the Raman spectrum of the pristine graphene sheet is plotted in Figure 4d, where the 'G' and '2D' peak intensities have been utilized as a reference to quantify the total number of graphene layers.[111] Following, in addition to the provided photocurrent responsivities for the monolayer graphene-based devices for a given concentration of gold NP solutions, the authors studied the tunneling behavior of the generated hot electrons in multilayered graphene-based tunneling junctions. As anticipated, the photocurrent responsivity values are decreased while the layer number increases, since fewer hot electrons were tunneled through the barrier (see Figure 4e). To deliver more insight on the plasmonic hot electron tunneling process, Du and teammates performed time-resolved differential reflection measurements. In this limit, the measured differential reflection signals ($\Delta R/R_0$) from the gold NPs for various cases are shown in Figure 4f. The associated delay time constants (refers to the lifetime of electrons in the excited state) of the examined samples are obtained as following: 0.76± 022 ps, 1.81±0.23 ps, and 2.28±0.32 ps, respectively. These results indicate that the induced hot electrons stay longer in multilayer graphene sheets rather than in a monolayer sheet, owing to the Landau damping effect and possible fabrication imperfections, such as wrinkles, during the additional graphene layer transfer step. Furthermore, a visualization for the hot electron generation process is given in Figure 4g. As illustrated, by thee absorption of incident beam and localization of surface plasmon resonance (LSPR), the surface plasmons are decayed either radiatively or non-radiatively, in which the hot electrons are generated through the latter one. On the other hand, when the graphene is injected with hot electrons, the Fermi level shifts upward *via* charge transfer operation controlled by the work functions of gold ($W_m$ ~ 4.7 eV) and graphene ($W_g$ ~ 4.5 eV) (Figure 4h).[65] Nevertheless, eventually, the aggregation of positive charges (holes) in gold NPs and negative charges (electrons) in graphene gives rise to an internal electrostatic field, which would change the direction of the charge transfer (Figure 4i). Besides, the light absorption by gold NPs induces a temperature increase within the system, which results the recombination of electrons and holes. The effect of incident laser power on the induced tunneling current is investigated in Figure 4i. For the lower values (<30 µW), the current increases linearly, but for the values higher than 30 µW, the power dependence of the current is changed.

As another study, Fang and collaborators[112] reported the fabrication and characterization of a graphene-enhanced plasmonic photodetector, consisting of Fano-resonant plasmonic antenna clusters sandwiched between two graphene monolayers (Figure 5a). In Figure 5b (left side), firstly, the authors fabricated the plasmonic antennas and the source-drain electrodes on a graphene sublayer. Subsequently, the system covered by another graphene layer to constitute the proposed sandwich formation (Figure 5b (right side)).[113] Besides, for the device provided in Figure 5b, the antennas (dimers, heptamers, and nonamers) were only fabricated in areas 1, 3, and 5, respectively, whereas the remaining regions (2 and 4) were left blank. To understand where the strongest antenna-graphene coupling occurs, Raman imaging is conducted with 785 nm laser illumination. Considerable Raman intensity enhancements are obtained in the regions of 3 and 5, since the heptamer and nonamer antennas are resonant with this wavelength (Figure 5b, inset). Next, the authors examine the morphology of the additional graphene sheet located on the heptamer array by the SEM image. As can be seen in Figure 5c, the top graphene monolayer covers the cluster profile and relaxing through the gaps between each heptamer. Moreover, the band-crossing energy or Dirac point of the utilized graphene is observed at a gate voltage ($V_G$) of ~30 V (Figure 5d), by applying a source-drain voltage of 1 mV. It is significant to note that the device mobility is calculated as $\mu = \left[\frac{dI_{DS}}{dV_G}\right] \times [L/(WC_iV_{DS})]$,[114] where $L$ is the channel length (~50 µm), $W$ is the channel width (~10 µm), and $C_i$ is the real capacitance per unit area between the channel and the back-gate (~1.2 × 10-8 F/cm$^2$). The estimated mobilities vary from 350 to 1300 cm$^2$ V$^{-1}$s$^{-1}$. Fang et al.[112] also measured the local photocurrent within the device by conducting a line scan of a laser (785 nm wavelength, 1 µm beam spot) between the drain and source electrodes, and an antisymmetric photocurrent response is observed (Figure 5e). The inset of Figure 5e illustrates the zero current point and the line scan path. One can obviously perceive that the antenna including regions boost the photocurrent, in comparison to the other regions. In the dimer antenna limit, the induced photocurrent possesses a highly polarization-dependent response (Figure 5f). On the other hand, the heptamer antennas (under 785 nm illumination) exhibit much larger photocurrent values, almost three times larger than that of the dimer structure and eight times larger than the pristine graphene regions of the device. This is because of large absorption cross-section, high field-enhancement, and boosted yield of hot electrons features of the heptamer antenna. Additionally, a gate bias (from -40 V to +40 V) is applied to actively control the Fermi level of the graphene layer. The measured photocurrents for the blank and heptamer antenna patterned region indicate analogous characteristics: the photocurrent goes to zero when the laser focus shifts to the center of the graphene channel. For a larger negative gate bias, the maximum photocurrent is obtained at a position near the drain and source electrodes. Conversely, when the gate bias is positive, the measured photocurrent downs to zero. Any further increase in the $V_G$ modifies the magnitude and the direction of the photocurrent. One can mention that all of these features are affected by the local variations in the band filling of graphene layer for different $V_G$. In Figure 5h, the blue dashed line represents the Fermi level of graphene, whereas the solid blue line indicates the Dirac point of graphene. It is significant to note that the titanium contact dopes graphene, due to its relatively low work function.[115] Thus, the Fermi level within the region of electrodes will be incremented slightly toward the Dirac point. On the other hand, the energy difference ($\Delta\phi$) between the doped graphene band and Fermi level forms a Schottky barrier, which is independent from the applied bias. The bias only alters the Fermi level of the graphene located in the middle of each area of the device. Since the gold NPs are deposited without requiring titanium adhesion layer, they would not inject any impurities to the graphene layer. As demonstrated in Figure 5h, the difference in Fermi level between the graphene channel and the doped graphene results to band bending around the electrode region. Within this area, the photocurrent reaches its maximum for the steepest band bending (at $V_G$=-40 V), owing to internally generated electric field. For the positive bias voltages, the Fermi level of the graphene channel rises and the band bending become less important. Particularly, when $V_G$=+20 V, the band is almost flat and the induced photocurrent is trivial. By increasing the bias voltage, the band starts to bend in the opposite direction and the inner electric field becomes larger. Since a gate voltage of 20 V is needed to have a flat band condition, the photocurrent obtained for +40 V is lower than that obtained for -40 V (Figure 5g).

## 4. Free carrier absorption (FCA) of silicon

Heavily doped silicon strongly absorbs infrared beam,[116] while intrinsic silicon is highly transmissive to this bandwidth.[117] In doped-silicon photodetectors, technically, three major absorption mechanisms can be considered:[116,118] 1) bandgap or intrinsic absorption, 2) impurity level-to-band absorption, and 3) free carrier absorption (FCA). The intrinsic absorption constitutes that the energy of the incoming photon must be higher than the bandgap of the semiconductor, while the impurity level-to-band absorption has been described for extrinsic infrared photodetectors. In the FCA limit, the photon energy is absorbed by free carriers either in conduction or valance band. All these characteristics can be seen in p-type silicon, as plotted in Figures 6a-6c. Theoretically, the FCA coefficient in semiconductors can be written as a function of the operating wavelength ($\lambda$), concentration of free EHPs ($\rho$), refractive index ($n$), mobility ($\mu$), and effective mass ($m^*$) as following:[119]

$$\alpha = \frac{q^3 \lambda^2 \rho}{4\pi^2 \varepsilon_0 c^3 n m^{*2} \mu} \quad (1)$$

where $c$ is the speed of light in a vacuum, and $\varepsilon_0$ is the permittivity of vacuum. Replacement of the already defined and constant numerical values for the corresponding parameters in the formula above, thus, the FCA coefficient can be rewritten as: $\alpha = 1.54 \times 10^{-17} \lambda^2 \rho / (\mu (m^*/m)^2)$. Correspondingly, the transmission coefficient for a silicon substrate with a given thickness ($d$) can be obtained by applying the absorption coefficient above as:[117,120]

$$T = \frac{(1-R)^2 e^{-\alpha d}}{1 - R^2 e^{-2\alpha d}} \quad (2)$$

where *R* is the reflection coefficient of a semi-infinite sample. As a leading experimental study, Schmid *et al.*[117] demonstrated the absorption spectrum of heavily doped silicon (n- and p-doped) at low and high temperatures for varying concentrations. Figures 6d and 6e exhibit the optical absorption coefficient as a function of photon energy for Arsenic (As)- and Boron (B)-doped silicon at 4 K and 300 K for different concentrations, respectively. In the n-type Arsenic-doped silicon, the FCA coefficient hugely increases for 24 to 40 $\times 10^{18}$ cm$^{-3}$, while in the p-type Boron-doped silicon, explicitly, the excessive FCA is obtained in the specimens with the carrier density of 70 to 120 $\times 10^{18}$ cm$^{-3}$.

Traditionally, intrinsic silicon-based photodetection subwavelength platforms suffer from transparency to the photons with energies below its bandgap (~1.12 eV), which makes this semiconductor not proper for NIR applications. This limitation was addressed by either heavily doping silicon or employing graphene in the design of silicon-based devices. More precisely, heavily doped silicon strongly absorbs infrared, mid-infrared, far-infrared, and terahertz lights using its FCA feature, which is an intraband transition of carriers within its conduction or valance band.[116,121] Strictly speaking, Boron-doped p-type silicon absorbs 100% of the incident infrared beam. As a promising technique, an appropriate integration of plasmonic gratings with low-doped p-type silicon (~$10^{15}$ cm$^{-3}$) has been employed to enhance the responsivity and photocurrent of NIR photodetectors by reducing the height of the Schottky barrier.[122] Although there have been significant advances in augmenting the performance of infrared photodetectors, the quality of these devices still needs to be enhanced. Very recently, Tanzid *et al.*[56] utilized a combination of hot-carrier generation and FCA to boost the NIR photodetector photon-yield. This was successfully accomplished by employing both generated hot-carriers in plasmonic gratings and FCA in heavily doped silicon. In this work, the researchers studied and compared the plasmonic response of Au and palladium (Pd) both experimentally and numerically. Here, we only summarize and demonstrate the photodetection response of the developed device for gold gratings. The photodetector mechanism and SEM image are demonstrated in Figures 7a-7c, in which the heavily doped p-type substrate absorbs the decayed hot electrons from the metallic grating. In addition, the inherent FCA changes the p-type silicon carrier mobility. The operating mechanism is clearly demonstrated in Figure 7a, where the transition from heavy hole to the light hole level gives rise to a smaller effective mass.[123] These variations in the hole effective mass leads to a higher carrier mobility in the doped silicon, resulting in a reduction of device resistance during light exposure. Consequently, FCA of p-type silicon amplifies the generation of the photocurrent by reducing the effective resistance of the entire plasmonic nanodevice. By employing accurately defined geometries for the metallic gratings (specified in the caption of Figure 7d), the unbiased photoresponse of the p-type silicon-mediated narrowband device consisting of 200 nm gold gratings on 2 nm titanium ($w$= 900 nm, $g$= 250 nm) is illustrated for two different doping densities ($5 \times 10^{18}$ cm$^{-3}$ (0.01 Ω.cm) and $6 \times 10^{16}$ cm$^{-3}$ (0.3 Ω.cm)) under longitudinally polarized beam illumination. At the peak response wavelength of 1375 nm, the responsivity of the highly doped p-type substrate is almost 4.5-fold greater than the responsivity on the lower doped p-type silicon substrate.

The photoresponsivity of the plasmonic device was further enhanced by applying bias across the system. The responsivity variations as a function of voltage for the device based on gold gratings at the peak position of the resonance is plotted in Figure 7e. For the following geometries: $w$= 900 nm, $h$= 200 nm, $g$= 250 nm, the device shows a continuous increasing responsivity as bias increases. Specifically, a larger than 1 A/ W photoresponsivity was achieved at 1375 nm wavelength for gold structures at a small applied bias of 275 mV. As a key parameters, the researchers analyzed the plasmonic and photoresponse of the device as a function of grating width ($w$), while the gap between gaps are fixed to 250 nm. These results are demonstrated in Figure 7f, where the unbiased responsivities of the fabricated detectors were measured. In this regime, the responsivity extremes monotonically red-shifted towards the longer wavelengths with increasing grating width. It is strongly claimed that the obtained responsivity was 3 to 5 times stronger than the analogous narrowband photodetectors.[94] Such a unique combination of photoexcited hot carriers and FCA coefficient in p-type silicon substrate leads to giant enhancement in the photoresponsivity and performance of the infrared plasmonic photodetector.

## 5. Enhanced photoresponsivity using toroidal resonances

As explicitly elucidated in the previous sections of this Review, the plasmonic photodetectors facilitate the transformation of incident photons into electrical signals through hot electron generation *via* plasmon decay. This causes a huge interest to the plasmonic photodetection platforms owing to their unique capabilities, such as focusing the incoming electromagnetic energy down to subwavelength regimes.[124] Over the last couple of years, advanced photodetection mechanisms based on plasmonic antennas and atomically thin layers have been studied to produce highly efficient hot electrons.[125-127] In particular, one can generate more hot electrons by exciting dark resonances (e.g. Fano resonance) through well-engineered geometries,[128-131] due to the suppressed scattering cross-section. Despite of the recent advances to improve the performance of optoelectronic devices, such as utilization of free-carrier absorption in doped-materials,[132] the power conversion efficiency and field confinement ability of photodetectors and the resulting photocurrent still need to be enhanced.

As an independent member of the electromagnetic multipole family, toroidal moments have been introduced.[133-136] Owing to the inherently weak far-field radiation signature of these moments in comparison to the classical electromagnetic modes, the detection of these moments is complicated.[137] Besides, one need to apply Lorentz and Feld-Tai theorems to analyze the induced electromagnetic fields of these time-dependent multipoles.[138] Recently, researchers have been developed artificially engineered metadevices to excite prominent toroidal multipoles from terahertz to optical frequencies.[139-141] Based on the toroidization phenomena, the dynamic toroidal dipole moment can be strongly focused in a tiny spot, which give rise to an exceptional electromagnetic field confinement.[142-144] At the resonant wavelength point, the toroidal structure preserves the energy of generated charge oscillations effectively because of the reduced far field emission rates, and this feature would be beneficial to substantially amplify the performance of the hot electron-based photodetection technology.

In this context, Ahmadivand *et al.*[73] have developed a new class of plasmonic photodetector by utilizing extremely squeezed and robust charge-current configurations, arising from judiciously designed periodic toroidal-resonant Au arrays on top of a doped Si substrate. [More information on the design and spectral response of the proposed plasmonic meta-atom is available in Ref. 73] As indicated in Figure 8a, oppositely aligned magnetic moments (**m**) are produced at the peripheral curve-shaped pixels under y-polarized incident light. The mismatch between the flux direction of magnetic charges and the resulting currents form an oblique charge-current configuration (known as toroidal dipole) (see the artistic schematic in Figure 8a, where the direction of the induced surface currents and the toroidal dipole moment is illustrated). The induced displacement surface current density across the entire nanoplatform at the toroidal dipole resonance is also provided. Next, the authors verified the excitation of toroidal dipolar mode decomposition analysis, in which the power of the scattered multipoles emitting from the scatterer are plotted. Based on the results shown in Figure 8b, one can understand that the main contributors of this profile analysis are toroidal dipole (**T**), in conjunction with electric dipole (**p**) and magnetic dipole (**m**), electric quadrupole ($\mathbf{Q^{(e)}}$), magnetic quadrupole ($\mathbf{Q^{(m)}}$), and electric octupole ($\mathbf{O^{(e)}}$). In Figure 8c, theoretical cross-sectional charge-current map is plotted to visualize the formation of the head-to-tail charge-current configuration across the meta-atom (see the dashed region). Furthermore, a 3D E-field density map is demonstrated in Figure 8d to indicate the importance of the capacitive gaps between the central and peripheral resonators, and how the intensity of the E-field is strengthened within the capacitive gaps. Following, with the help of these remarkable features, the authors reveal the possibility of hot electron generation using the proposed toroidal plasmonic metadevice by applying NIR illumination and bias. In Figure 8e, an artistic diagram for the devised photodetection mechanism is presented, where a series of standalone structures and drain/source electrodes are featured. Similar to Figure 8d, maximum field confinements are achieved at the gap regions. A cross-sectional snapshot of the corresponding E-field map and the induced surface current distribution (J) are also included (see Figure 8f). Besides, the top-view SEM image of the fabricated device is exhibited in Figure 8g. The operation mechanism of the studied photodetector can be explained by the stimulation of substantially intense and confined plasmons along the metallic unit cells (which yields strong absorption) and free carrier absorption of the doped Si (particularly, in p-type Si). At the specific toroidal dipole wavelength, generation of the dynamic hot electrons and resulting photocurrent enhancement occur, because of the extreme field localization and substantial absorption cross-section. Moreover, due to the lessened electron-electron scattering in the toroidal meta-device, the number of excited electrons transferred to the doped-substrate is increased.[145] With this way, the transition and accumulation of photoexcited electrons are expedited before they recombined. In continuation, the mechanism behind the hot electron production and photocurrent generation is investigated. To this end, the proposed device is fabricated in two different doping regimes, as n- and p-doped, and the induced photocurrent is both numerically calculated and experimentally measured (see Figures 8h and 8i, respectively). As expected, the device possesses extremely high photocurrent response in the p-doped Si regime,[118] due to augmented carrier mobility of Si substrate, longer lifetime of the photogenerated carriers (~100 μs), and generated large amount of electrons at the Schottky interface. It is significate to mention that the carrier concentration in both doping regimes is set to $2 \times 10^{19}$ cm$^{-3}$. By applying a gate bias in the range of 0 mV-500 mV, the photoexcited electrons and holes are drifted to the forward- and reverse-biased electrodes, respectively. In Figure 8j, the numerically calculated photoresponsivity response of the metadevice is depicted in both limits. Similar to the photocurrent and absorption plots, a distinct photoresponsivity response is obtained: ~14.5 mA W$^{-1}$ and ~29 mA W$^{-1}$ for the n-type and p-type device, respectively. Furthermore, as a strategic parameter of the photodetection concept, IQE (which is the ratio between the total number of charge carriers leading to the induced photocurrent ($I_p$) and the total number of photons absorbed by the structure) of the toroidal nanoplasmonic platform is calculated as:[59] IQE = $(I_p/q)/(S_{abs}/h\upsilon)$, where $q$ is the elementary charge, $S_{abs}$ is the absorbed optical power contributing to the photocurrent, $h$ is the Planck's constant, and $\upsilon$ is the frequency of the impinging light. As demonstrated in Figure 8k, an IQE of 38.5% is achieved for the p-type toroidal metadevice, whilst this value drops to ~30% for the n-type photodetector, and one can enhance this parameter further by incorporating the provided toroidal nanoplatform with absorptive 2D layers. In addition, the authors numerically quantified the minimum noise equivalent power (NEP) and detectivity ($D^*$) of the toroidal NIR photodetector in Figures 8l and 8m, respectively. The minimum NEP of the metadevice is obtained as 5.4 pW Hz$^{-1/2}$ based on the following:[146] $i_n = \left(\frac{4k_BT\Delta f}{R}\right)^{1/2}$, in which $k_B$ is Boltzmann's constant, $T$ is the room temperature, $\Delta f$ is the frequency bandwidth, and $R$ is the resistance of the photodetector. In addition, the maximum detectivity value of the toroidal metadevice is calculated as 7.06 × 10$^9$ Jones, using the following formula:[147] $D^* = \mathcal{R}(\Delta)^{1/2}/(i_n^2 + i_{nb}^2)^{1/2}$, where $\mathcal{R}$ is the photoresponsivity and $i_{nb}$ is the noise current owing to the background radiation.

**Conclusions and outlook**

This Review highlights the recent advances in photocurrent generation at IR wavelengths using plasmonic platforms. It is shown that the well-engineered nanoplasmonic structures can significantly boost light absorption and convert incident photons into carriers with extra degree of control over their energies. Although several efficient and innovative approaches for tailoring high-responsive light sensing tools have been discussed and summarized in this Review, there are still number of ongoing novel methodologies to optimize these processes for the next-generation nanophotonic technologies. In principle, there are numerous parameters that define the performance of an infrared photodetection device, such as carrier relaxation and lifetime, FCA, hot carrier energy, etc. Taking advantage of these specifications, various high-photon yield, narrowband, wavelength-selective, and linearly functional plasmonic light sensing devices have been developed successfully. This enables designing advanced tools for bio-imaging, telecommunication systems, imaging, military, and astronomy applications. On the topic of plasmonic-induced carrier generation, by bridging the fundamentals of light absorption in semiconductors and electron or hole decay from metallic components, we provided a deep insight into the yield and energy distribution of carriers. It is shown that how the inherent FCA in doped semiconductors can be combined with plasmonically decayed carriers in IR regime towards huge photocurrent generation. This was achieved by considering the fact that plasmons are inherently extremely fast excitations, and by coupling hot carrier generation on these nanostructures to the excitations in proximal systems, therefore, the timescale can be monitored.

Advances in the concept of light to current conversion have led to the rise of novel and promising plasmonic photodetectors based on unconventionally dark-hot resonances.[73] In this area, toroidal photodetectors have been introduced as a new direction in tailoring high-responsive IR devices. Large photocurrent, fast operation, significant linear dynamic range, and low noises make these platforms as reliable and potential alternatives for classical plasmonic photodetectors. The ability to localize the electromagnetic field in a single spot and provide a huge absorption cross-section allows for the generation of huge amount of carriers with low losses due to scattering of electrons. It is also experimentally verified that a judicious combination of toroidal meta-atoms and doped semiconductors leads to substantially large photocurrent and exquisitely high-photon yield. Thus, this concept can be considered as an alternative route to revitalize the plasmonic response of integrated IR photodetectors.

**Figures**

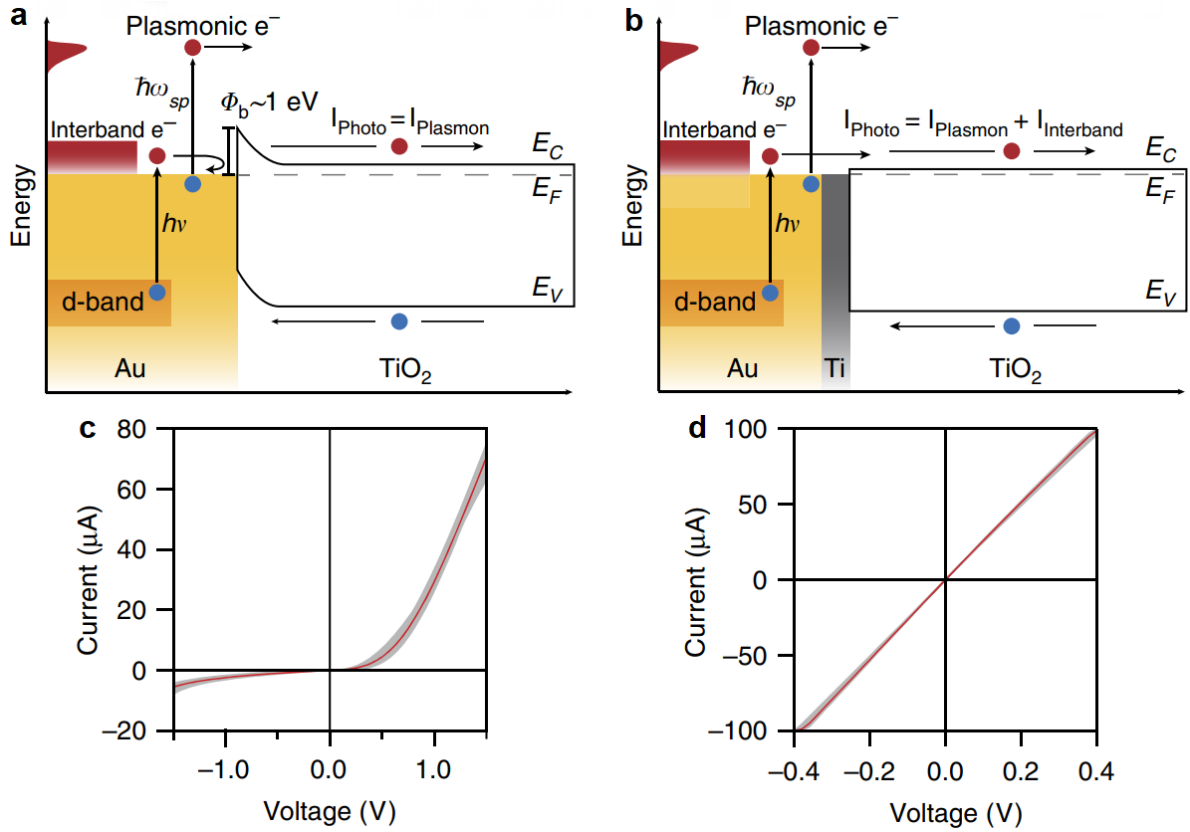

**Figure 1.** (a) A gold-TiO$_2$ Schottky and (b) a gold-Ti-TiO$_2$ Ohmic device. Carrier generation by direct photoexcitation results from the excitation of d-band electrons, into the conduction band. Their low energy prevents them from crossing the Schottky barrier. Ohmic devices have no effective barrier and allows for collection of carriers created by this method. Current-voltage (I–V) curves of Schottky (c) and Ohmic devices (d).[76] Copyright 2015, Nature Publishing Group.

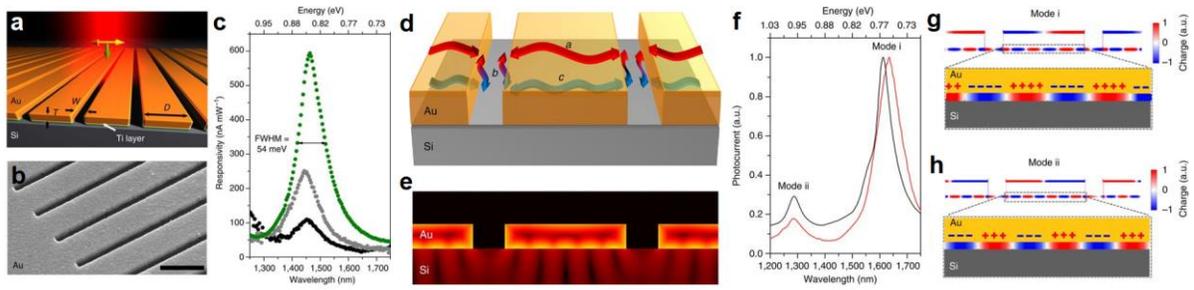

**Figure 2.** (a) Schematic of the proposed device as gold grating on an n-type silicon substrate with a thin layer of titanium adhesion layer. (b) SEM image of gold grating with *T*=200 nm, D=950 nm, and W=250 nm. The scale bar is 1 µm. (c) The responsivity plot for three different gold grating thicknesses as T=93 nm (black), 170 nm (grey), and 200 nm (green). (d) Possible forms of the induced surface plasmons as '*a*': oscillating at the top surface of the grating, '*b*': oscillating through the slits, and '*c*': oscillating at the bottom surface of the grating (or at the Schottky interface). (e) Calculated plasmonic heat distribution. As indicated, most of the hot electron generation appears at the bottom surface of the gold layer. (f) Calculated (black) and experimental (red) responsivity curves for the grating structure with *T*=200 nm, *D*= 1100 nm, and *W*=250 nm. In both cases, the spectra has two modes as 'Mode i' and 'Mode ii'. Surface charge plot distribution at the interface for (g) 'Mode i' and (h) 'Mode ii', respectively.[94] Copyright 2013, Nature Publishing Group.

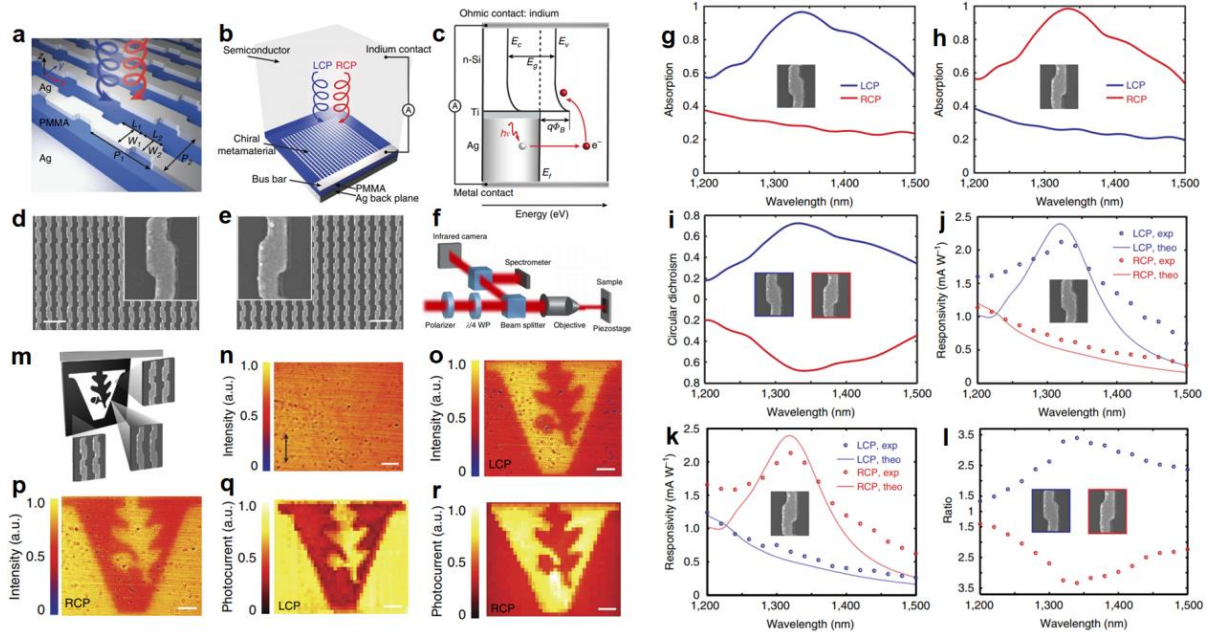

**Figure 3.** (a) Illustration of the chiral metamaterial. The dimensions of each metamolecules are following: $L_1$=125 nm, $L_2$=105 nm, $W_1$=115 nm, $W_2$=85 nm, $P_1$=335 nm, and $P_2$=235 nm. The thickness of the metamolecules, metal backplane, and the PMMA layer are 40 nm, 100 nm, and 160 nm, respectively. (b) Schematic of the proposed CPL photodetector. In this design, the circuit is completed through the silver bus bar and indium, which is utilized to form an Ohmic contact. (c) Energy band diagram of the photodetector. As described, the Schottky barrier is built between silicon and the adhesion layer. The photo-induced hot electrons within the silver metamaterial are transferred to the silicon over this interface. SEM images of the (d) LH and (e) RH metamaterial. The inset indicates a standalone version of the metamolecule (Scale bar: 500 nm). (f) Schematic of the utilized experimental setup. Experimentally measured optical absorption spectra under RCP (red) and LCP (blue) illumination for (g) LH and (h) RH metadevices. (i) Experimentally measured CD spectra for both RH (red) and LH (blue) metamaterials. Experimentally measured (dots) and theoretically calculated (solid line) photoresponsivity spectra under RCP (red) and LCP (blue) irradiation for (j) LH and (k) RH metamaterials. (l) Photocurrent polarization discrimination ratio spectra of RH and LH metamaterials. (m) Illustration of the university logo pattern, in which the black and white regions are filled with LH and RH metamaterials, respectively. Camera images of the metamolecule under (n) linearly polarized light, (o) LCP, and (p) RCP illumination. Scanning photocurrent maps of the metamaterial under (q) LCP and (r) RCP light. Scale bar, 10 μm.[97] Copyright 2015, Nature Publishing Group.

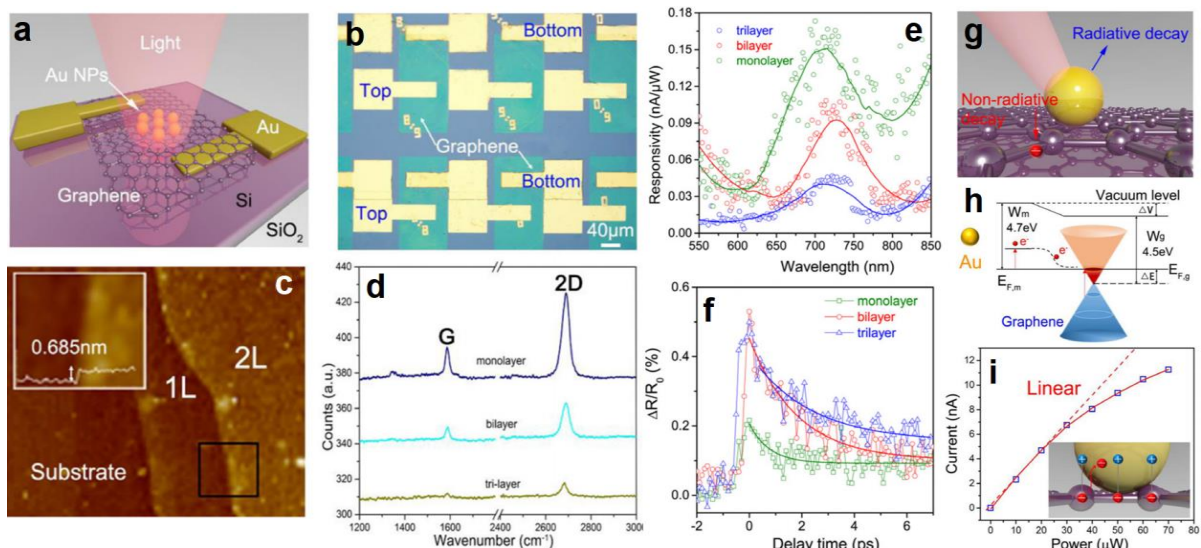

**Figure 4.** (a) Schematic of the vertical graphene-based photodetector. The incident light is focused between the gold electrodes, which includes gold NPs. (b) Optical microscopy image of the fabricated device under 50x objective lens. (c) Tapping-mode AFM measured topology results of monolayer and bilayer graphene sheets. (d) The Raman spectra of the pristine graphene. (e) The I-λ characteristics of the monolayer (green), bilayer (red), and trilayer (blue) graphene samples. The provided curve for the trilayer graphene layer is zoomed two times. (f) Delay time dependence of $\Delta R/R_0$ for gold NPs-monolayer, gold NPs-bilayer, and gold NPs-trilayer excited with light of 400 nm (with a pump power of 70 µW). The solid lines represent single-exponential fitting results. (g) An artistic illustration of the hot electron generation in a single gold NP by illumination. (h) Energy diagram of the graphene Dirac cone with gold NP. As a result of the illumination, electrons from the occupied energy levels are excited above the Fermi level with energies high enough to move into the Fermi surface. (i) A plot to demonstrate the dependence of the tunnel current on the incident laser power for a monolayer graphene. Inset: a schematic to represent the hot electron recombination because of the internal electrostatic field.[110] Copyright 2017, WILEY-VCH.

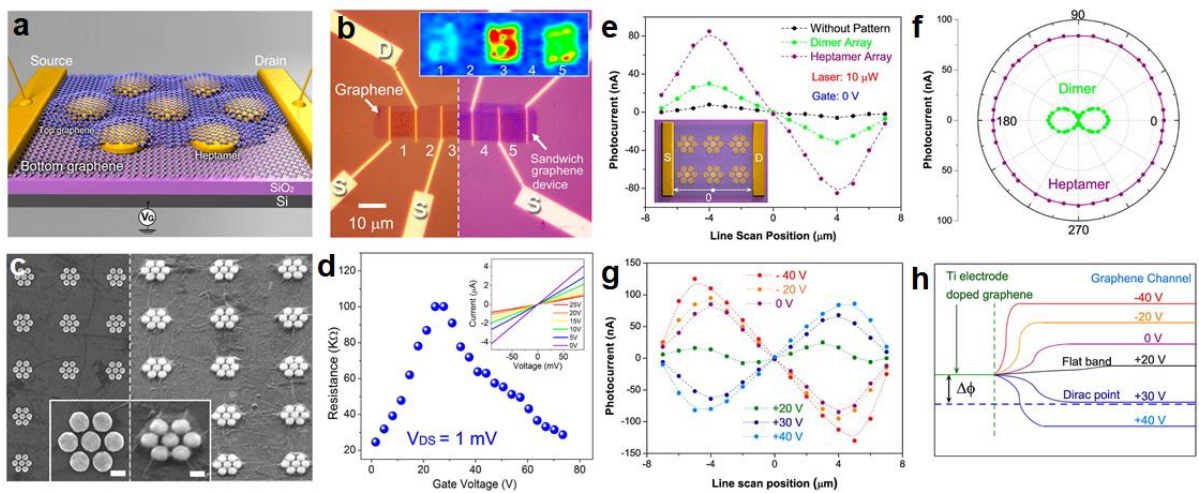

**Figure 5.** (a) Schematic illustration of the proposed graphene-antenna sandwich photodetector. $V_G$ is the gate voltage used to electrostatically dope the graphene. (b) Optical microscopy image of the fabricated device before (left) and after (right) deposition of the additional graphene layer. Inset: Raman mapping of the 'G mode' of device regions 1-5 under 785 nm illumination. (c) SEM image of the heptamer array fabricated in area '3'. (d) Electrical transport characteristics at a drain bias of 1 mV. Inset: *I-V* plots for different $V_G$ values (from 0 V to 25 V). (e) Photocurrent characterization of the sandwich devices. As plotted, the measurements indicate antisymmetric photocurrent responses from various regions of the device, acquired along the line scan direction. The black dots show the photocurrent values detected without depositing plasmonic antennas in the specific regions of the device. Inset: The zero point is depicted. (f) Polarization dependence of the generated photocurrent of antennas in region '1' (green dots) and region '3' (purple dots). (g) Measured photocurrent for $V_G$ varying between -40 V and +40 V in region '3', with the incident laser power of 10 µW. (h) A plot to illustrate the surface potentials concluded from the line scan experiments for each gate bias.[112] Copyright 2012, American Chemical Society.

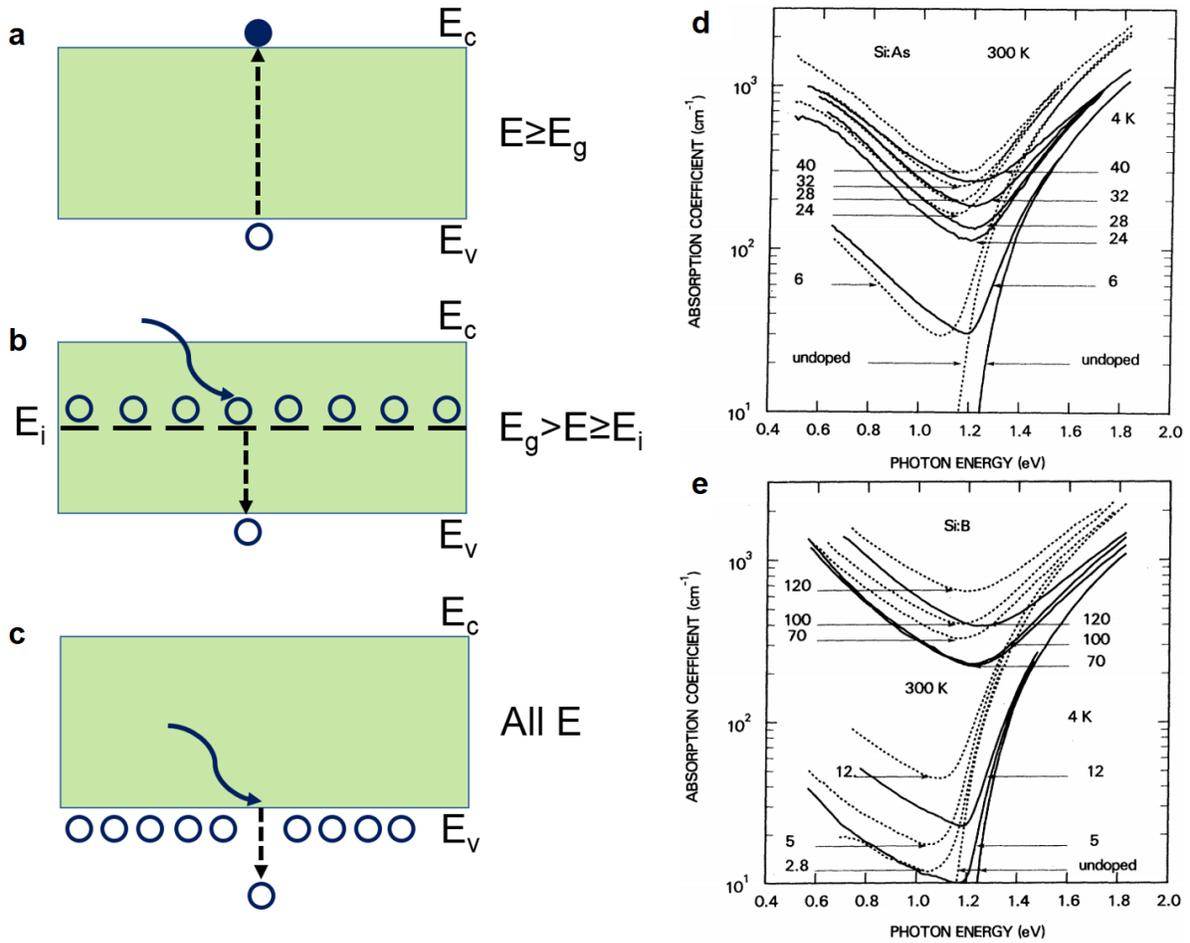

**Figure 6.** (a) Intrinsic, (b) extrinsic, (c) FCA in p-type semiconductors. Optical absorption coefficient as a function of photon energy for (d) Arsenic-doped and (e) Boron-doped silicon at 4 K and 300 K. The curves are labeled by their respective donor density divided by $10^{18}$ cm$^{-3}$.[117] Copyright 1981, American Physical Society

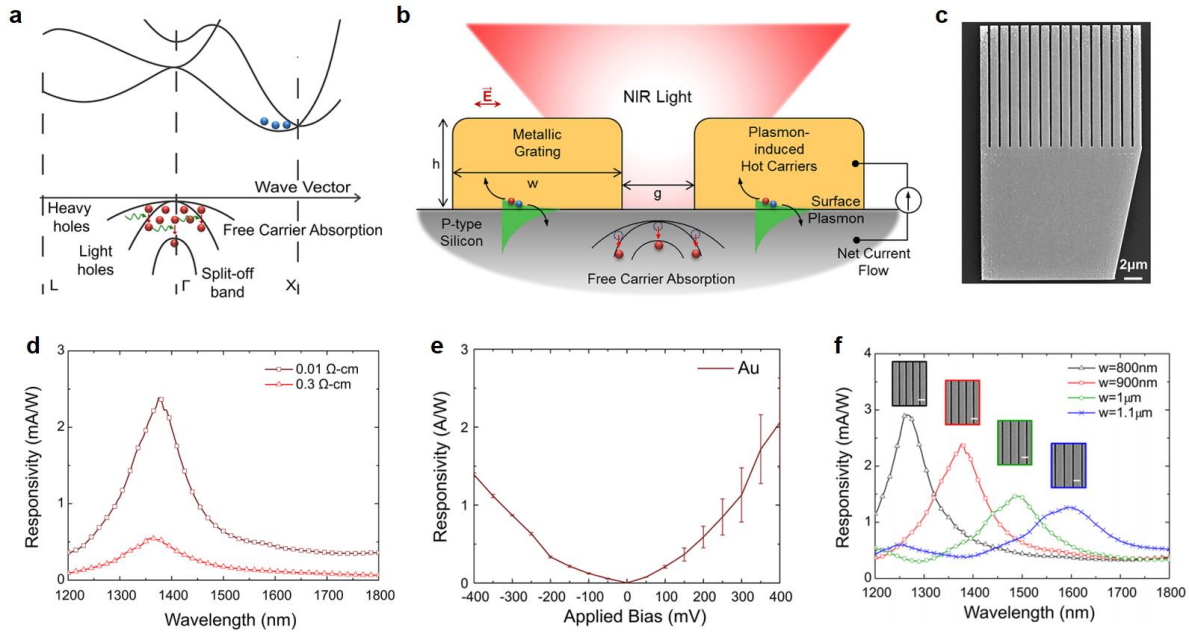

**Figure 7.** (a) Band diagram schematic of silicon showing the direct intraband transitions in the valence band corresponding to FCA. (b) Schematic illustration of a plasmonic photodetector that generates photocurrent by using hot carriers from metallic grating FCA in heavily doped p-type silicon substrate. The dimensions of the metallic grating are superimposed by $h$, $w$, and $g$. (c) SEM image of a gold grating nanostructure with a metallic contact pad on a silicon substrate. (d) Measured photocurrent responsivities of gold gratings ($w$= 900 nm, $h$= 200 nm, $g$= 250 nm) on p-type silicon substrate with 0.01 Ω·cm (doping concentration 5× $10^{18}$ cm$^{-3}$) and 0.3 Ω.cm (doping concentration 6× $10^{16}$ cm$^{-3}$) resistivities. (e) Measured responsivity of the same photodetector on 0.01 Ω·cm p-type silicon substrate with applied external bias across the device at λ= 1375 nm. (f) Measured responsivity of gold gratings ($h$= 200 nm, $g$= 250 nm) on p-type silicon substrate with 5× $10^{18}$ cm$^{-3}$ doping concentration. The insets show SEM images of the gratings in each case (scale bar =1 μm).[56] Copyright 2018, American Chemical Society

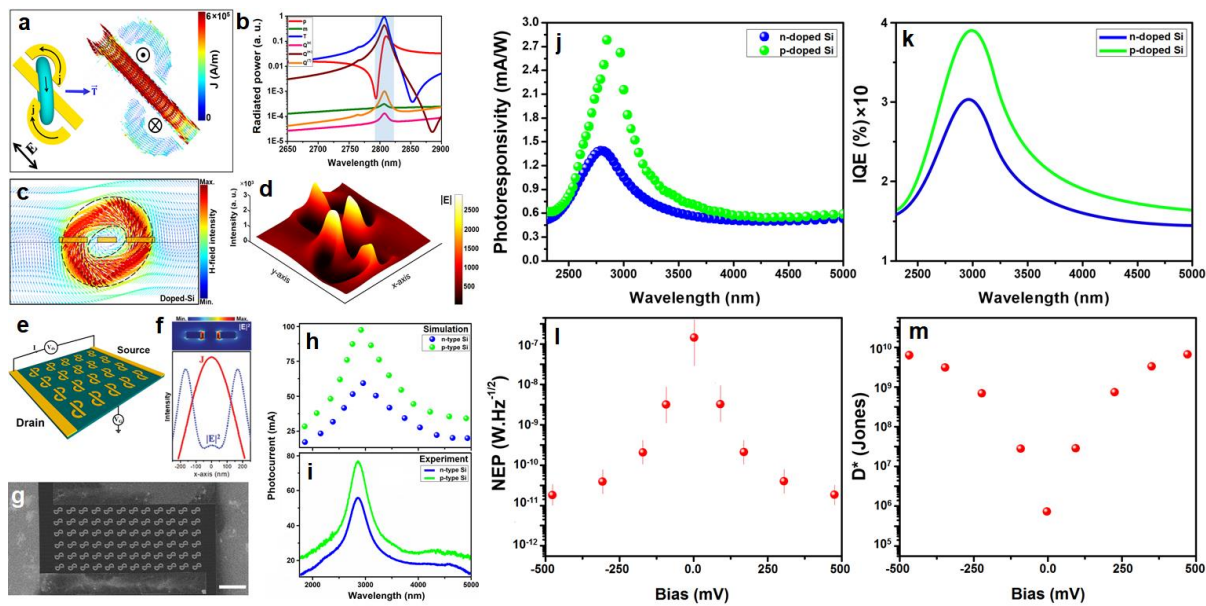

**Figure 8.** (a) The displacement surface current density across the entire platform under *y*-polarized light. (b) Calculated radiated power of each electromagnetic multipoles excited within the metadevice. (c) *xz*-plane of the induced H-field to verify the spinning charge-current configuration across the central bar resonator. (d) 3D E-field enhancement plot to show the extreme field confinement within the gaps. (e) Artistic view of the proposed toroidal photodetector. (f) Cross-sectional absolute E-field confinement and associated charge density (J) maps of the studied meta-atom. The latter indicates that the maximum J occurs at the central bar. (g) SEM image of the fabricated toroidal metamaterial (scale bar: 1µm). (h) Calculated and (i) measured magnetoelectric currents at the electrodes for both n- and p-type substrates (Here, the drain-source voltage is fixed to ±5 mV). (j) Photoresponsivity and (k) IQE performance of the device as a function of incidence for both n- and p-type substrates. (l) Calculated NEP and (m) photodetectivity of the toroidal photodetector at the toroidal dipole wavelength (2850 nm).[73] Copyright 2019, Royal Society of Chemistry.